\documentclass[prd,nofootinbib,twocolumn,superscriptaddress]{revtex4}

\usepackage{graphicx}
\usepackage{amsmath}
\usepackage{amsfonts}
\usepackage{amssymb}
\usepackage{dsfont}
\usepackage{dcolumn}
\usepackage{bm}
\usepackage{pstricks}
\usepackage{color}

\newcommand{\beqn}{\begin{eqnarray}}
\newcommand{\eeqn}{\end{eqnarray}}
\newcommand{\be}{\begin{equation}}
\newcommand{\ee}{\end{equation}}
\def \BB{B_s \to {\mu}^{+} {\mu}^{-}} 
\def \SI{\sigma_{\rm SI}(\chi p)} 

\begin{document}

\title{
Low Mass Neutralino Dark Matter in the MSSM   \\
with Constraints from $\BB$ and Higgs Search Limits}
\author{Daniel Feldman}
\affiliation{Michigan Center for Theoretical Physics,
University of Michigan, Ann Arbor, MI 48109, USA}

\author{Zuowei Liu}
\affiliation{C.N.\ Yang Institute for Theoretical Physics, 
Stony Brook University, Stony Brook, NY 11794, USA
}

\author{Pran Nath}
\affiliation{Department of Physics, Northeastern University,
 Boston, MA 02115, USA}


\preprint{YITP-SB-10-05; MCTP-10-04; NUB-3265 }


\begin{abstract}
The region of low neutralino masses as low as (5-10) GeV has attracted 
 attention  recently due to the possibility of excess  events above background
in dark matter  detectors.
An analysis of spin independent neutralino-proton cross sections  $\SI$
that
includes this low mass region is given. The analysis is done  in MSSM
with radiative electroweak symmetry breaking (REWSB).
It is found that cross sections as large as $10^{-40}$ cm$^2$ can be accommodated
in MSSM within the REWSB framework. 
However,  inclusion of sparticle mass limits from current experiments, as well
as lower limits on the Higgs searches from the Tevatron, and the current experimental upper limit
on $\BB$ 
significantly limit the allowed parameter  space reducing $\SI$ to lie below $\sim 10^{-41}$cm$^2$ or
 even lower for neutralino masses around 10 GeV.
These cross sections are  an order of magnitude lower
 than the cross sections needed to explain the reported data in the  recent  dark matter 
 experiments in the low neutralino mass region.
 \end{abstract}

\maketitle


\section{ Introduction} 
Recent experimental results on the direct detection of dark matter  \cite{Ahmed:2008eu,xenon,Lebedenko:2008gb} 
have made very significant progress  in increasing the sensitivity of detectors  to probe the spin 
independent scattering cross sections of WIMPs off target nuclei.
Specifically, the most recent five tower CDMS II result \cite{newcdms}
has reached the sensitivity of $3.8\times 10^{-44}$ cm$^2$ at a mass of 70 GeV. These results
have received attention in supersymmmetric models of cold dark matter \cite{cdms}.

More recently CoGeNT \cite{Aalseth:2010vx}
has reported results that  show some overlap with the parameter space where DAMA 
sees an excess of events \cite{Bernabei:2008yi}. In the context of the minimal supersymmetric standard model (MSSM), 
with electroweak symmetry  broken radiatively,
it is interesting to ask whether neutralino dark matter can give rise to 
detectable events  in the region where   CoGeNT/DAMA see an excess.
The reported excess occurs   in the low mass region  with neutralino mass as low as 
5-10 GeV. 
Indeed  light neutralinos  with masses $O(10) \rm GeV$ 
have  been entertained by several authors within the MSSM framework\cite{Bottino:2002ry,Bottino:2008mf,Bottino:2009km,Dreiner} and light dark matter using different frameworks has also been investigated\cite{Gunion,Feng:2008dz,Savage:2008er}.

In this brief note we investigate the neutralino low mass region in MSSM 
under the  Tevatron constraints from Higgs searches and on $\BB$. For the analysis here we choose the
framework of radiative breaking of the  electroweak symmetry with the aim of finding the parameter
 space of soft breaking that can generate spin independent neutralino proton
cross sections which can  lie in the range $(10^{-40}- 10^{-41})$ cm$^2$  needed
to get compatibility with the data in the low neutralino mass experiments. 
The large spin
independent neutralino proton cross sections of size $O(10^{-40})$ cm$^2$ arise via t-channel
Higgs exchange for large $\tan\beta$ which is typically required to be as large as 50 or larger. 
[Squark exchange in the s-channel can also produce competitive results if the 
 lower limit on the masses  is relaxed  below the current experimental limits 
 as will be evident in the discussion given in Sec.(II).]
However, the large $\tan\beta$ region is precisely the region where the $\BB$  becomes large\cite{gaur}.
It is known that the $\BB$ branching ratio 
 has a  $\tan\beta$ dependence, which grows as powers of $\tan\beta$ (the growth can
vary from  $\tan^2\beta$ to $\tan^6\beta$ depending on the part of the parameter space one is
in). For this reason the $\BB$ experimental limits produce a very strong constraint in regions of the parameter
space where $\tan\beta$ gets large which is what is needed to get a large spin independent neutralino 
proton cross section. We give now details of the analysis including the constraints from mass limits\cite{Feldman:2007zn} 
and from the $\BB$ branching ratio.

  \begin{figure}[t]
  \begin{center}
\includegraphics[width=8cm,height=6.5cm]{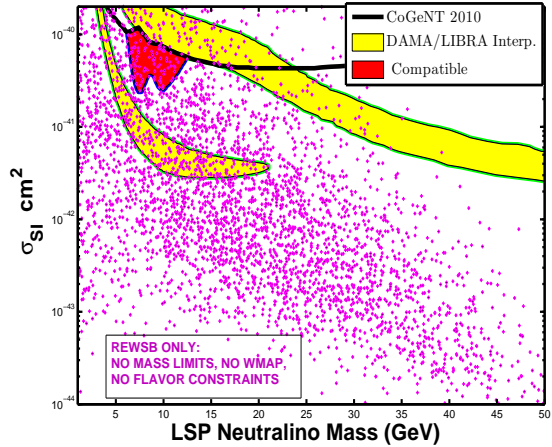}
\caption{(Color online) A display of the spin independent neutralino-proton cross section
$\SI$  as a function of LSP mass
with inclusion of REWSB constraints. 
The sparticle mass limits, WMAP constraints and flavor constraints
are not imposed. The analysis of this figure should be compared 
with the analysis of Fig(\ref{fig2}) which shows the large
reduction in the parameter space when sparticle mass 
and WMAP  constraint are imposed (see Bottino et al \cite{Bottino:2009km}).}
 \label{fig1}
  \end{center}
\end{figure}

 \begin{figure}[t]
  \begin{center}
\includegraphics[width=8cm,height=6.5cm]{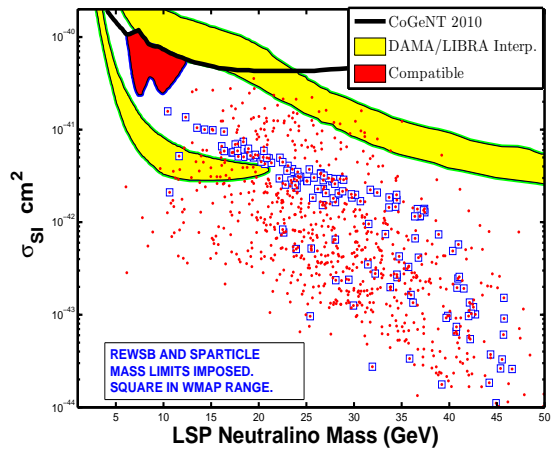}
\caption{(Color online)
A display of the spin independent neutralino-proton cross section
$\SI$ in MSSM  with inclusion of REWSB and mass limit constraints.
 The tagged boxed points accommodate the WMAP results (see text)
 while the untagged ones do not.  
Shown also are the CoGeNT limits\cite{Aalseth:2010vx}.}
 \label{fig2}
  \end{center}
\end{figure}
 \begin{figure*}[t]
  \begin{center}
\includegraphics[width=8cm,height=6.5cm]{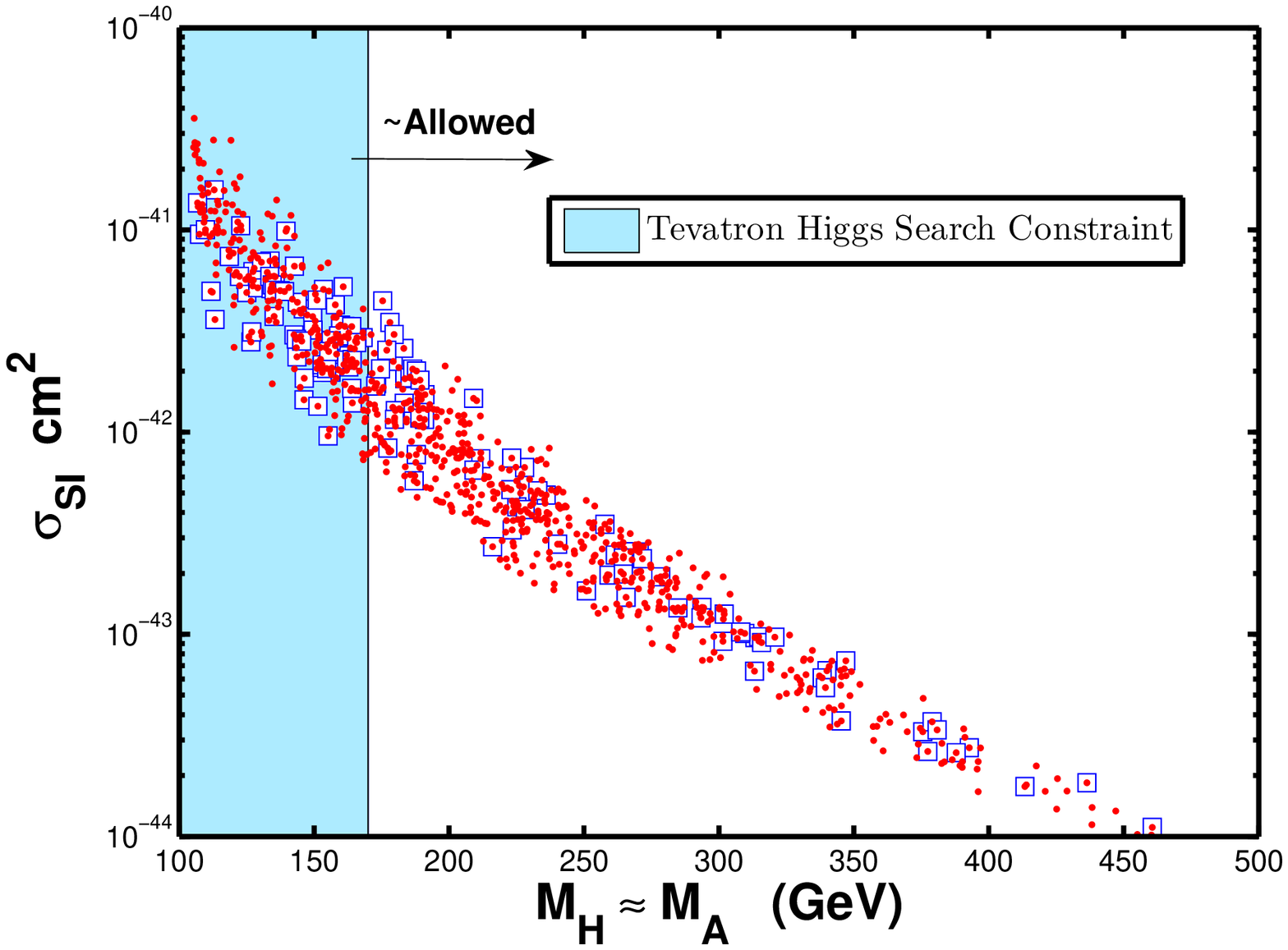}
\includegraphics[width=8cm,height=6.5cm]{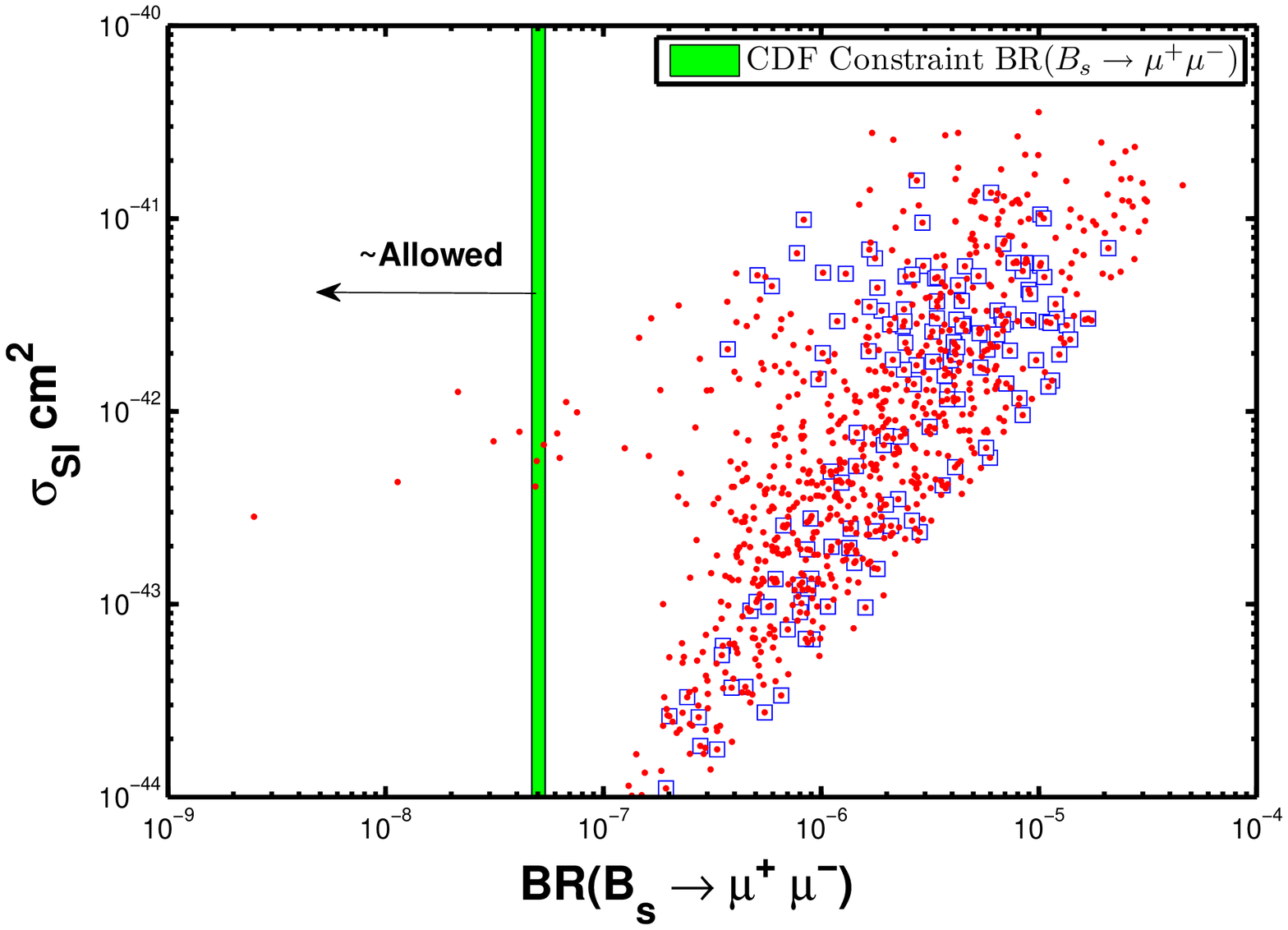}
\caption{(Color online) Left:  
An exhibition of the spin independent neutralino-proton cross section $\SI$ in MSSM 
as function of the CP even Higgs mass $M_H$.
The blue band is the approximate range of Higgs mass disallowed
by Tevatron searches. The analysis shows that the reach of the spin independent cross section is strongly
constrained by the mass of the Higgses \cite{Abazov:2008hu,Feldman:2007fq}. 
Right:
Spin independent cross section correlated to the branching ratio $\BB$.  
The region of maximal $\SI$ is inconsistent with constraints from CDF data on $\BB$\cite{Aaltonen}. {The models shown
have large $\tan \beta$} and light scalars {(see left panel and  Sec. {\ref{XX}}  for the full parameter space)} }.
 \label{fig3}
  \end{center}
\end{figure*}

\section{An MSSM Analysis with REWSB \label{XX}}
The excess reported in 
   DAMA and CoGeNT experiments requires  a relatively large $\SI$, i.e., 
   \beqn
\SI \sim 10^{-40}~{\rm cm^2}, ~~~~~~ m_{\rm WIMP} \sim (5-10) ~\rm GeV.
\eeqn
We discuss now the possibility whether cross sections of the above size can arise in MSSM.
The elastic spin independent cross section, $\SI$, in the MSSM 
is controlled dominantly by the Higgs exchange and by the squark exchange depending on the relative
lightness of the scalars.
For the Higgs  or the squark exchange diagrams 
the cross section is maximized for large $\tan\beta$ and $\SI$ scales as \cite{Chattopadhyay:1998wb}
\beqn 
 \SI_{\rm max} \sim \frac{(g_Y g_2)^2 |n_{11} {n_{1,\left[4,3\right]}}|^2}{4 \pi M^2_W} \frac{[1,\tan^2 \beta]}{m^4_{ S}} F_p  ,\\  
\left[S = h, (S= \tilde q ~or ~H)\right].
\label{eq1}
\eeqn
Here $(n_{11 = \tilde B},n_{12 = \tilde W},n_{13= {\tilde H}_1},n_{14= \tilde{H}_2})$ stand for the
bino, wino, and Higgsino components of the neutralino
and $F_p$  depends on the neutralino couplings to the proton. 
The sensitivity of $\SI$ to MSSM parameters enters essentially via the Higgs/squark masses
and via the eigencomponents $n_{11}$, $n_{13}$ etc of  the neutralino wave function. 
In the analysis of radiative breaking of the electroweak symmetry we assume the
MSSM parameter space with non-universalities in the gaugino masses.
Eq.(2) is an approximate form and in the analysis we have used the full form
for $\SI$. Thus
implementing micrOMEGAs and SuSpect \cite{Belanger:2008sj}
we scan the parameter space consistent with REWSB and with GUT scale unification in the ranges
$m_0 \leq$ 500 GeV, $m_{1/2} \leq$ 200 GeV,  $|A_0/m_0| \leq  4$, 
the latter guided by constraints on 
charge and color breaking, with $\tan \beta \in (45,60)$. We take the
sign of $\mu$ to be positive and take the top quark mass to be 171 GeV.  In the gaugino
sector we assume ${ M}_a/m_{1/2}=(1+\Delta_a)$, for $a=1,2,3$ where $\Delta_1 \in (-1,0)$
and  $\Delta_{2,3} \in (0,1)$ where the ranges are guided in part by the restriction
that the lightest chargino $\chi^{\pm}$ mass should be greater than 100 GeV. 
Beyond
this, we require the lightest CP even Higgs to have a mass of 100 GeV or larger, and
we require the squark and gluino mass to be larger than 300 GeV.
To account for theoretical uncertainties in the large $\tan \beta$ region
 we take a broad range  around the  WMAP \cite{Komatsu:2010fb} value of the neutralino
 relic density so that  $\Omega h^2 \in [0.08,0.15]$.
 
An analysis of the $\SI$ under the constraints of radiative electroweak symmetry breaking
 but without the  imposition of any sparticle mass limits is given in Fig.(\ref{fig1}).
Here one finds that the allowed parameter space does encompass the regions where
CoGeNT and DAMA/LIBRA see excess events.
 However, in the analysis of Fig.(\ref{fig1}) we have 
not imposed any sparticle mass limits or  the  $\BB$ branching ratio constraint. 
 In Fig.(\ref{fig2}) we give an analysis of $\SI$ as a function of the neutralino mass
 with inclusion of the experimental sparticle mass limits.  
 The points within the boxes  are  those that satisfy the relic density while the small circles 
do not.
One finds that the maximal $\SI$  in the allowed parameter space is an order of magnitude below the
CoGeNT region while some of the  points in the allowed parameter space  do lie in the DAMA/LIBRA region. 
Next, in the left panel of  Fig.(\ref{fig3}) we give a plot of $\SI$ vs  the CP even Higgs mass $M_H$. 
Here we exhibit the parameter space  constrained  by the Tevatron data on the
Higgs searches. Inclusion of this  constraint reduces the size of $\SI$ to be no
greater than {$ \sim 5 \times 10^{-42}$ cm$^2$}.
 We note that  in the analysis of both  Fig.(\ref{fig2}) and the left panel of Fig.(\ref{fig3}) 
we have not included the $\BB$ constraint. We will see below that inclusion of the $\BB$ constraint further reduces
the maximally allowed size of $\SI$.

The CDF data give the following upper limit  from the process  $\BB$ \cite{Aaltonen}
\beqn
{\mathcal Br}( B_s \to \mu^{+}\mu^{-})   < (5.8/4.7) \times 10^{-8}~~~( 95/90 \% {\rm CL}).
\eeqn
We will see that these limits are  
 quite constraining  at large $\tan \beta$  specifically for regions where the spin independent cross
 sections get large. 
$\BB$ has a leading $\tan^6 \beta$ dependence  which is further modified by loop corrections
\cite{gaur,Buras:2002wq}  so that 
\beqn
\BB \sim  1.92\times 10^{-6}(\tan \beta)^6 (M_A/{\rm GeV})^{-4} 
\nonumber\\
\times \frac{(16 \pi^2\epsilon_Y)^2}{(1+ (\epsilon_0 + y^2_t \epsilon_Y)   \tan \beta)^2 (1+ \epsilon_0 \tan \beta)^2}~~
\label{5}
\eeqn
where   $\epsilon_0$ and $\epsilon_Y$ are loop corrections
(Eq.(\ref{5}) is an approximate  form for $\BB$ and in the analysis we have used the full form.).
Eq.(\ref{5}) exhibits the fact that a light SUSY Higgs boson and a large $\tan\beta$ tend to enhance the $\BB$ 
branching ratio which is subject to strong Tevatron limits.  
These Tevatron constraints 
 are exhibited in the right panel of Fig.(\ref{fig3}).  Here the allowed region of the parameter
space consistent with the upper limit on $\BB$  does not fall within the region
consistent with the region where CoGeNT or DAMA are sensitive to $\SI$.  
 As is easily seen very few parameter points lie in this region consistent with the $\BB$
 constraint and
 relatively large $\SI$, 
and those that fall in the region consistent the $\BB$ constraint have $\SI$ which does not exceed {$\sim 5  \times 10^{-42}$ cm$^2$}. 

{The above result is found when fixing  
$\Sigma_{\pi N}=55$~MeV with 
$f^{p}_{d}=0.033$, $f^{p}_{u}=0.023$, $f^{p=n}_{s}=0.26$,
$f^{n}_{d}=0.042$, $f^{n}_{u}=0.018$,
with the hadronic  matrix elements 
$\langle N| m_q\overline{\psi}_q \psi_q|N\rangle=f^N_{q}M_N.$
Of the above quantities, the largest uncertainties 
 in the calculation of  $\SI$ arise from variations in the
pion-sigma nucleon term, followed by  uncertainties in the 
strange quark contribution (for a recent analysis see \cite{Savage}).
It is useful to illustrate the effects
of these uncertainties.
 In the region of large $\tan \beta$, 
(the region of interest here where the $\SI$ can be enhanced)
for the case of a light neutralino with mass of 7 GeV
we find this can lead to an uncertainty as large as a factor of $\sim 3$
in the $\SI$
over the window $\Sigma_{\pi N}=64 \pm 8 \rm ~MeV$.
The uncertainty in $f^p_{s}$ can generally lead to a factor of $\sim 2$.
Taking into account the above, coupled with sparticle mass constraints
and the WMAP constraint, we do not find any models that can fall in the
Cogent preferred region.
 It should be noted that the very recent lattice QCD simulations from one group \cite{Giedt}
may imply a further uncertainty (which could even lower the cross section).
Further detailed analyses  on such simulations need to  be carried out.}

 We note 
that extending the parameter space of the  models to include larger scalar masses will suppress  $B_s \to \mu^{+}\mu^{-}$ 
and of course weaken this constraint. However, heavy scalars  will reduce the $\SI$ and it is for this reason
we have focused on the parameter space above. We discuss this further below.

\section{Discussion and Extensions}
{ We now discuss our findings further by extending the set of models. 
Thus it is also interesting to  extend the   parameter space to include
a larger set of non-universalities. As a check on our conclusions, we 
have carried out an exhaustive scan  of  high scale supergravity models with non-universalities 
in various sectors. These include non-universalities in the gaugino  (NUG) masses,
non-universalities in the Higgs boson (NUH) masses, non-universalities in the 
third generation scalar (NU3) masses, and non-universalities in the trilinear couplings
(NUA) in the third generation sector $A_{s}$, $s=\tilde \tau, \tilde b, \tilde t$.
We utilize  Monte Carlo simulations generating  {several million} 
candidate models for the non-universal cases discussed above.
Thus in total we investigate NUG, NUA, NUH, and NU3
simulating a total of 4 million candidate models.
Specifically the parameter space considered in the Monte Carlo scans 
are as follows: 
$
0 <  m_0/{\rm GeV} < 1000 ,
1 < \tan\beta < 60,
169 < m_t/{\rm GeV}  < 173,
0 < M_1/{\rm GeV} < 1000,
0 < M_2 /{\rm GeV} < 1000,
0 < M_3 /{\rm GeV} < 1000,
-5 <  A_{\tau}/m_0 < 5,
-5 < A_{t}/m_0 < 5,
-5 < A_{b}/m_0 < 5,
0 <  M_{H_u}/{\rm GeV} < 1000,
0 < M_{H_d}/{\rm GeV} < 1000.
$
}

{Within this rather large parameter space
of high scale models and under the constraint of REWSB and other experimental constraints already discussed}, our conclusions remain that of Sec.(\ref{XX}).
{We note in passing that in the region of interest  the $\mu$ parameter typically lies in the range 100 -300 GeV and  the smaller $\mu$'s are associated  
with significant Higgsino components of the neutralino which enhance the $\sigma_{SI}$. However,
$\sigma_{SI}$ significantly larger than $10^{-42}$ cm$^2$ still appear difficult to achieve even with the inclusion of 
non-universalities.}
{Further, } {one finds that while the 
REWSB alone allows an LSP neutralino as low as even a  GeV,  the present
mass bounds on the supersymmetric particles, along with the $B_s \to \mu^{+}\mu^{-}$ and WMAP constraints
 allow  only a 20 GeV LSP in the models
discussed in Sec.(\ref{XX}). 
 Inclusion of  a lower limit on the $\tilde \tau$ mass of 98.8 GeV 
 pushes the lower limit on the neutralino mass to about 30 GeV.  Finally, we remark  that for the LSP neutralino mass in the (20-40) GeV region,
in SUGRA models with REWSB, the light neutralino masses are  further constrained by  direct detection experiments \cite{Feldman:2007fq}.
This constraint was not discussed in the detailed analysis of Ref. \cite{Dreiner} which focused on collider bounds and cosmological bounds
of light neutralinos. 
}  

\section{Conclusion}
Recent results by CoGeNT and DAMA are sensitive to light neutralino masses in the $(5-10) \rm GeV$ region
with spin independent cross sections in the region of $10^{-40} \rm cm^2$. This relative largeness
of the spin independent cross section, $\SI$,  can be accommodated  in the MSSM in the framework of 
radiative electroweak symmetry breaking. Inclusion of sparticle mass limits from current experiments, as well
as lower limits on the Higgs searches from the Tevatron, and the current experimental upper limits
on  the $\BB$ branching ratio,
significantly reduce the allowed parameter  space. The residual parameter  space with neutralino
{masses greater than 10 GeV  has} a drastically reduced $\SI$  which is 
not in excess of {$\sim 10^{-41}$ cm$^2$} and is significantly lower than the $\SI$ needed in the
low neutralino mass regions of experiments presently probing low mass dark matter. {Inclusion of
all known constraints, both collider, satellite, and terrestrial direct detection constraints 
implies that a 10 GeV or lighter neutralino cannot accommodate the CoGeNT  preferred region in models with 
only the MSSM
spectra.}

\noindent
{\em Acknowledgments}:  
This research is  supported in part by DOE grant  DE-FG02-95ER40899 (Michigan Center for Theoretical Phyics, (MCTP)),  
and NSF grants PHY-0653342 (Yang Institute for Theoretical Physics, (YITP), Stony Brook), and  PHY-0757959 (Northeastern University). 

\end{document}